%% file: SECRYPT.tex
\documentclass[a4paper,twoside]{article}

\usepackage{epsfig}
\usepackage{calc}
\usepackage{amssymb}
\usepackage{amstext}
\usepackage{amsthm}
\usepackage{amsmath}
\usepackage{multicol}
\usepackage{pslatex}
\usepackage{apalike}
\usepackage{SCITEPRESS}     
\usepackage{mathtools}
\usepackage{booktabs}
\usepackage{epstopdf}

\usepackage{multirow}
\usepackage{hhline}
\usepackage{ifpdf}
\usepackage{subfig}
\usepackage{makecell}
\usepackage[T1]{fontenc}
\usepackage{mathptmx}
\DeclareGraphicsExtensions{.png,.jpeg,.pdf}
\graphicspath{{png/}{jpg/}{pdf/}}

\renewcommand{\vec}[1]{\mathbf{#1}}
\begin{document}

\title{On The Instability of Sensor Orientation in Gait Verification on Mobile Phone}

\author{\authorname{Thang Hoang\sup{1}, Deokjai Choi\sup{2} and Thuc Nguyen\sup{3}}
\affiliation{\sup{1}Faculty of Information Technology, Saigon Technology University, Ho Chi Minh City, Vietnam}
\affiliation{\sup{2}Department of Electronics and Computer Engineering, Chonnam National University, Gwangju, South Korea}
\affiliation{\sup{3}Faculty of Information Technology, University of Science VNU-HCMC, Ho Chi Minh City, Vietnam}
\email{thang.hoangminh@stu.edu.vn, dchoi@jnu.ac.kr, ndthuc@fit.hcmus.edu.vn}
}


\abstract{Authentication schemes using tokens or biometric modalities have been proposed to ameliorate the security strength on mobile devices. However, the existing approaches are obtrusive since the user is required to perform explicit gestures in order to be authenticated. While the gait signal captured by inertial sensors is understood to be a reliable profile for effective implicit authentication, recent studies have been conducted in ideal conditions and might therefore be inapplicable in the real mobile context. Particularly, the acquiring sensor is always fixed to a specific position and orientation. This paper mainly focuses on addressing the instability of sensor's orientation which mostly happens in the reality. A flexible solution taking advantages of available sensors on mobile devices which can help to handle this problem is presented. Moreover, a novel gait recognition method utilizes statistical analysis and supervised learning to adapt itself to the instability of the biometric gait under various circumstances is also proposed. By adopting PCA+SVM to construct the gait model, the proposed method outperformed other state-of-the-art studies, with an equal error rate of 2.45\% and accuracy rate of 99.14\% in terms of the verification and identification aspects being achieved, respectively.}

\onecolumn \maketitle \normalsize \vfill

\input{p1-introduction.tex}
\input{p2-relatedwork.tex}
\input{p3-instabilityproblem.tex}
\input{p4-proposedmethod.tex}
\input{p5-evaluation.tex}
\input{p6-conclusion.tex}

\bibliographystyle{apalike}
{\small
\bibliography{example}}

\end{document}

%% file: p1-introduction.tex
\section{\uppercase{Introduction}}
\label{sec:introduction}

\noindent Over recent years, mobile devices have greatly evolved from primitive machines for voice and text communication to personal intelligent assistants and are becoming more familiar to everybody. A survey of the mobile market forecasted that mobile subscriptions will reach 9.3 billion by 2019, 5.6 billion of which will be for smart phones. Mobile facilities include not only making calls or sending text messages, but also cover a variety of utilities such as data storage, entertainment and Internet transactions. Since such devices are likely to be portable personal computers, sensitive personal data is accumulated in them, which might lead to supplemental security demands. Additionally, mobile devices tend to become increasingly miniaturized and light, which makes them a lot easier to lose. A huge amount of personal data could become exposed to criminals. The most popular authentication technique currently used in mobiles still relies on the traditional method of using a secret code, such as a PIN, visual pattern or password \cite{Breitinger10}.  These techniques are not highly effective considering the problems of memory and security \cite{Breitinger10}. Alternatives using biometric traits, such as the face, fingerprint or on-line signature, have been introduced recently on mobiles, which have helped to mitigate the limitations of password-based methods \cite{Jain04}. However, all of these methods strongly rely on user cooperation and might therefore be annoying and obtrusive in frequent use. Users are forced to pay attention and perform explicit gestures in order to be authenticated. 

Besides, the number of mobile applications is exploding these days and these various applications might require different levels of security. Indeed, a trade-off between usability and security needs to be taken into consideration. For instance, retrieving the user's daily schedule does not require the same level of security as making an Internet banking transaction. Applying the same verification scheme to all applications requiring different levels of security would be somewhat cumbersome. Thus, it is necessary to provide miscellaneous authentication mechanisms on the mobile adapting to different security level requirements to optimize the user-device interaction. Accordingly, an implicit authentication technique needs to be investigated, which aims to enhance the user experience and ameliorate mobile security. Human gait has been studied for a long time and shown to be as an effective behavioral biometric trait \cite{Jain04,Fish93,Whittle03}. Identification using gait signals captured by wearable sensors has been introduced recently and has achieved positive results \cite{Ailisto05,Gafurov09}. Verification on mobiles leveraging gait characteristic of individuals has significant advantages in terms of user friendliness and security, in comparison to other biometric modalities \cite{Mjaaland11,Derawi13}. Specifically, gait signals can be implicitly captured while the user is walking without his or her intervention. From the security perspective, it is difficult to counterfeit authentic gait patterns even if the impostor could record the walking style of the genuine user \cite{Mjaaland11}. Conversely, a copy of a fingerprint or face could be easily obtained and the system security fully depends on the resistance of the sensor. However, in most existing gait recognition systems using wearable sensors, the acquiring sensors are likely to be fixed in a specific orientation and position, such as the waist, ankle or hip, to ensure that the shape of the acquired gait signals is similar \cite{Derawi13,Ailisto05,Gafurov09,Derawi101,Gafurov10}. It can be seen that these positions might be inappropriate, especially in the mobile context. Moreover, fixing the orientation of the device seems impossible in practice. 

In this paper, we propose a novel gait recognition scheme which can be used for user verification or identification on mobile device that can adapt to the actual usage in reality. We pay attention to the context that the mobile is placed in the front pocket, which is the most appropriate location for the device in daily use \cite{Breitinger10}. This study mainly focuses on addressing the instability problem of sensor's orientation that frequently arises when the device is flexibly attached with its owner in practice. Furthermore, gait is likely to be considered as a behavioral biometric which is not as robust as other physiological traits since it is affected by many physical and environmental conditions, such as the clothing, footwear, ground material, mood, health, age, weight, etc. Therefore, applying pattern matching, as in recent studies \cite{Derawi101,Derawi102,Derawi13,Gafurov09,Gafurov10,Rong07}, to deal with all these circumstances could be inefficient. What is more, since the mobile is generally carried and accessed by its owner, gait signals can be captured frequently and continuously. We prefer to leverage machine learning techniques to adapt to the variation of the gait characteristics over time. Any change in the gait patterns can be implicitly labeled and notified to the system to update the outdated model when the system frequently fails to verify  the user.

In summary, our main contributions are:

\begin{itemize}
\item[--] Addressing the instability of sensor's orientation when gait signal of individual is captured. A simple but effective solution for this issue taking advantage of the available sensors in mobile devices is presented (Section \ref{sec:instabilityproblem}). 

\item [--]Proposing a gait recognition model using statistical analysis and supervised machine learning (Section \ref{sec:gaitrecognition}). The results achieved in our experiment show that the proposed system has lower error rates, in comparison to other state-of-the-art methods (Section \ref{sec:experiment}).
\end{itemize}

%% file: p2-relatedwork.tex
\section{\uppercase{Related Works}}

\noindent Human gait data are considered to represent the particular style and manner in which human feet move and, hence, contain information of identification. On a more detailed level, the mechanism of human gait involves synchronization between the skeletal, neurological and muscular systems of the human body \cite{Fish93}. In 2005, H. Ailisto et al. were the first to propose gait verification using wearable sensors \cite{Ailisto05} and this area was further expanded by Gafurov et al. \cite{Gafurov09}. In general, sensors are attached to a particular position such as the ankle \cite{Gafurov09,Gafurov10,Li11,Terada11}, hip \cite{Gafurov09,Derawi101,Sprager09}, waist \cite{Ailisto05,Ngo14}, arm \cite{Gafurov09}, or multiple positions \cite{Pan09,Mondal12} on the body to record locomotion signals. The acquiring sensors can be gyroscopes or rotation sensors, but an accelerometer is most commonly used to capture gait signals. The most popular approach in this field is based on pattern matching, in which the gait signals are captured, preprocessed and then split into separate patterns. Various distance metrics such as the Dynamic Time Warping (DTW) \cite{Derawi101,Gafurov10,Rong07,Derawi102}, Euclidean distance \cite{Terada11}, auto-correlation \cite{Ailisto05}, and nearest neighbors \cite{Pan09} are used to calculate similarity scores between the given patterns and the stored templates. The second approach is based on machine learning. Feature vectors are extracted and supervised learning is adopted to construct general gait verification or identification models \cite{Lu14,Mondal12,Hoang13,Frank10}

Although wearable sensors have been implemented with a variety of success rates, they have some limitations. For example, those sensors are relatively expensive and cumbersome, due to their size and weight. The sensor interface is still under development. Recently, the improvement of micro electromechanical (MEM) technology has helped to miniaturize such sensors, thus allowing them to be integrated in mobile devices. Gait identification has thus been implemented on mobile devices \cite{Derawi13,Lu14} for the past few years. In comparison to wearable sensors, mobile sensors are designed to be cheaper, simpler and, as a result, their quality is not guaranteed. For instance, since the sampling rate is low and unstable, the noise level is rather high.  Derawi et al. \cite{Derawi101} demonstrated these deficiencies by re-implementing Holien et al.'s work \cite{Holien08}. The authors achieved an EER of 20.1\%, in comparison to the original EER of 12.9\%.

%% file: p3-instabilityproblem.tex
\section{\uppercase{The Instability of Sensor Orientation }}\label{sec:instabilityproblem}

\begin{figure}[!h]
\vspace{-0.2cm}
\centering
\includegraphics[width=7.5cm]{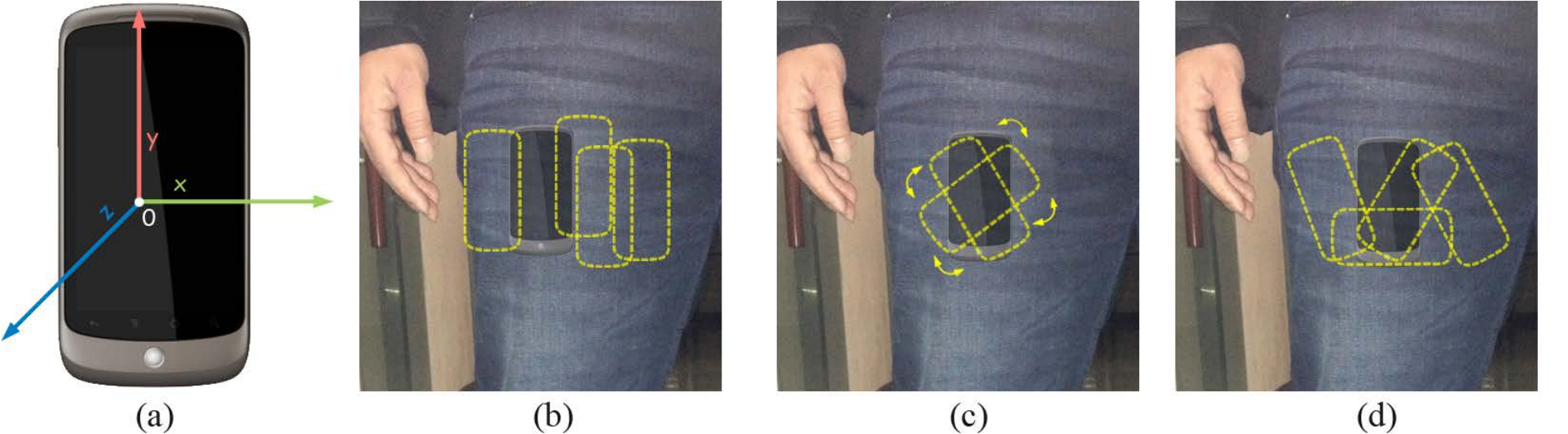}
\caption{(a) Mobile coordinate system, (b) misplacement error, (c) disorientation error and (d) disorientation errors and misplacement errors.}
\label{fig:disorientation}
\vspace{-0.1cm}
\end{figure}

\subsection{Problem statement}\label{sec:problemstatement}
Figures \ref{fig:disorientation}b--d illustrate the instability of the mobile in terms of its orientation and position when it is put freely in the pocket. Because walking is a slow activity with a moderate fluctuation, any strong acceleration is likely to last no longer than a few tenths of a second. Consequently, the impact of different positions in the pocket is not significant and is considered as noise. This can be mitigated by applying an effective noise filtering algorithm which will be described in Section \ref{sec:noiseelimination}.

Meanwhile, the instability of the mobile's orientation (namely the disorientation for short) significantly affects the quality of the acquired gait signals. Due to the design of the mobile accelerometer, wherein the gait signals are captured in 3 separate dimensions, the shape of the acquired signals fully depends on the relative orientation between the mobile and its carrier. So, the instability of the mobile orientation would make the gait signals in each separate dimension dissimilar. This obstacle could ruin the effectiveness of gait verification or identification systems. For instance, the accuracy rate of pattern matching approaches can be reduced when dealing with dissimilar gait signals. This circumstance will be illustrated in Section \ref{sec:impactdisorientation}. Furthermore, based on our observations, gait cycle-based segmentation can be easily performed on the gait signal in the dimension which represents the vertical walking direction (viz. the $Y-$dimension as in Figure \ref{fig:disorientation}a). Due to the disorientation problem, it is difficult to determine the correct dimension which reflects the vertical walking direction. Errors in the segmentation phase might propagate to subsequent processing phases, so that the effectiveness of the system can be compromised. Finally, extracting reliable features from dissimilar gait signals could be a problematic issue. 

Generally, the main objective of resolving the disorientation problem is to maintain the accuracy rate of the mobile gait verification or identification systems under practical conditions. A small part of this problem was solved in \cite{Hoang13}. However, there are unrealistic assumptions and constraints which could make the authors' proposed method difficult to apply in practice. We present a more flexible solution to this problem in the following section.

\subsection{Proposed solution}\label{sec:accelerationtransformation}
A simple but effective strategy to handle the disorientation is making gait signals always be represented in a fixed coordinate system which is insensitive to the device's orientation. In other words, acceleration vectors representing gait signals should be transformed from the instable mobile coordinate system  to a stable one. Based on the availability of sensors in the mobile, the Earth coordinate is likely to be considered as the effective fixed system to represent the collected acceleration samples. To do that, it is mandatory to collect various kinds of sensor data during the gait sensing period. The following section describes all necessary sensors need to be activated to collect enough data used in this study.

\subsubsection{Data acquisition}\label{sec:dataacquisition}
\noindent Obviously, the first sensor needs to be activated to capture gait signal is the mobile accelerometer. The accelerometer senses forces acting on the mobile in the three orthogonal axes of $ X, Y, Z $ (Figure \ref{fig:disorientation}a). A sequence of acceleration samples output by the accelerometer during walking is recognized as the gait signal. Each sample is a 3-dimensional vector, wherein each component is a combination of the forces of gravity and user motion acting on each dimension.
\begin{equation}\label{eq1}
\vec{a} =(a^{(X)},a^{(Y)},a^{(Z)}) ,	
\end{equation}
where $ a^{(D)} $ is the acceleration value sensed on the $ D- $axis of the mobile.

Due to the characteristics of the accelerometer, the raw acceleration samples always comprise gravitational acceleration components. In order to obtain samples which only involve pure gait signals of individuals, we eliminate the impact of gravity by additionally activating a virtual sensor of gravity to determine the gravitational acceleration components on the 3 axes of the mobile during the gait capture process. The output of the gravity sensor is a 3-component vector
\begin{equation}\label{eq2}
\vec{g}=(g^{(X)},g^{(Y)},g^{(Z)}),
\end{equation}
where $ g^{(D)} $ is the gravitational acceleration on the $ D- $axis of the mobile.

Furthermore to resolve the disorientation problem, we activate a synthetic sensor of orientation, along with the two sensors above to monitor the orientation states of the mobile. As in the case of the accelerometer and gravity sensor, the output of the orientation sensor is a 3-component vector 
\begin{equation}\label{eq2}
\vec{o} =(\alpha,\beta,\gamma)  ,
\end{equation}
where $ \alpha,\beta,\gamma $ represent the degrees of rotation around the $ Z-, X-, Y- $ axes of the mobile respectively.

Note that both orientation sensor and gravity sensor are all virtual sensors whose outputs are normally synthesized from two physical sensors: the accelerometer and the geomagnetic field sensor. These sensors are getting more and more popular, appearing in most modern smartphones so that all mandatory sensor data needed in this study can be easily obtained in practice.

\subsubsection{Gait signal transformation} 
Let us assume that after a gait sensing period, we obtain $ n $ vectors of acceleration $ \vec{a}_i $, orientation $ \vec{o}_i $ and gravity $ \vec{g}_i $
\begin{equation}\label{eq3}
\begin{aligned}
\vec{A}&=\begin{bmatrix}
\vec{a}_{1} & \hdots & \vec{a}_{i} & \hdots & \vec{a}_{n}
\end{bmatrix}^\top  \in \mathbb{R}^{n\times3}, \\
\vec{G}&=\begin{bmatrix}
\vec{g}_{1} & \hdots & \vec{g}_{i} & \hdots & \vec{g}_{n}
\end{bmatrix}^\top  \in \mathbb{R}^{n\times3}, \\
\vec{O}&=\begin{bmatrix}
\vec{o}_{1} & \hdots & \vec{o}_{i} & \hdots & \vec{o}_{n} 
\end{bmatrix}^\top\\ & = \begin{bmatrix}
\vec{\alpha}_{1} && \vec{\beta}_{1} && \vec{\gamma}_{1}\\ \vdots && \vdots && \vdots \\ \vec{\alpha}_{i} && \vec{\beta}_{i} && \vec{\gamma}_{i}\\ \vdots && \vdots && \vdots \\ \vec{\alpha}_{n} && \vec{\beta}_{n} && \vec{\gamma}_{n} 
\end{bmatrix} \in \mathbb{R}^{n\times3}.
\end{aligned}
\end{equation}

First, the influence of gravity on the acquired acceleration samples is eliminated to obtain the pure gait signal.
\begin{equation}\label{eq4}
\vec{A} \leftarrow \vec{A}-\vec{G}.
\end{equation}

For each rotation vector $ \vec{o}_i $ in $ \vec{O} $, we calculate a rotation matrix  $ \vec{R}_i $ which will be used to transform the acceleration vector in the mobile coordinate system to the Earth coordinate system.
\begin{equation}\label{eq5}
\noindent
\setlength{\arraycolsep}{1pt}
\resizebox{0.48\textwidth}{!}
{ $
\vec{R}_i = \begin{bmatrix}
\cos{\alpha}_i\cos{\gamma}_i-\sin{\alpha}_i\sin{\beta}_i\sin{\gamma}_i & \sin{\alpha}_i\cos{\beta}_i & \cos{\alpha}_i\sin{\gamma}_i+\sin{\alpha}_i\sin{\beta}_i\cos{\gamma}_i \\ -\sin{\alpha}_i\cos{\gamma}_i-\cos{\alpha}_i\sin{\beta}_i\sin{\gamma}_i & \cos{\alpha}_i\cos{\beta}_i & -\sin{\alpha}_i\sin{\gamma}_i+\cos{\alpha}_i\sin{\beta}_i\cos{\gamma}_i \\ -\cos{\beta}_i\sin{\gamma}_i &- \sin{\beta}_i & \cos{\beta}_i\cos{\gamma}_i
\end{bmatrix}
\\$
}
\end{equation}

Finally, we transform the gravity-free acceleration vector representing in the mobile coordinate system to the new fixed system by multiplying the vector with the corresponding rotation matrix.
\begin{equation}\label{eq5}
\vec{a}_i \leftarrow \vec{a}_i \vec{R}_i.
\end{equation}

The gait signal after transformation is denoted as
\begin{equation}
\setlength{\arraycolsep}{1pt}
\resizebox{0.42\textwidth}{!}
{$
\vec{A}= \begin{bmatrix}
\vec{a}_1 \\ \vdots \\ \vec{a}_i \\ \vdots \\ \vec{a}_n
\end{bmatrix} = \begin{bmatrix} {\vec{a}_1}^{(X)} & {\vec{a}_1}^{(Y)} & {\vec{a}_1}^{(Z)} \\ \vdots & \vdots & \vdots \\ {\vec{a}_i}^{(X)} & {\vec{a}_i}^{(Y)} & {\vec{a}_i}^{(Z)} \\ \vdots & \vdots & \vdots \\ \vec{a}_n^{(X)} & {\vec{a}_n}^{(Y)} & {\vec{a}_n}^{(Z)} \end{bmatrix} = \begin{bmatrix}
\vec{a}^{(X)} & \vec{a}^{(Y)} & \vec{a}^{(Z)}
\end{bmatrix}.
$}
\end{equation}

The acceleration vectors after transformation are presented in the Earth coordinate system, wherein the new $ Z- $dimension represents the vertical direction which is perpendicular to the ground, whereas the new $ X- $ and $ Y- $ dimensions represent the horizontal plane which is parallel to the ground. These transformed $ X- $ and $ Y- $dimensions always point towards the East and the magnetic North Pole respectively regardless of the walking direction. However, due to the fact that the user can walk in any direction in the horizontal plane, gait signals in the transformed $ X- $ and $ Y- $ dimensions which are captured from a session can be dissimilar to those captured from other sessions respectively. Therefore, instead of using the signals in each separate dimension of $ X $ and $ Y $, we utilize the combined signal of $ X-Y $
\begin{equation}
\vec{a}^{(XY)}  = (a_1^{(XY)}, \hdots , a_i^{(XY)}, \hdots ,a_n^{(XY)} ),
\end{equation}
where $ a_i^{(XY)}= \sqrt{{(a_i^{(X)})}^2+{(a_i^{(Y)})}^2}. $

In other words, the gait signals will be finally represented in the 2 dimensions of the Earth, wherein the transformed $ Z- $ and $ XY- $ axes represent the vertical and horizontal directions of walking, respectively. Moreover, the magnitude of the gait signal is additionally utilized as an additional dimension for gait representation.
\begin{equation}
\vec{a}^{(M)}  = (a_1^{(M)}, \hdots ,a_i^{(M)}, \hdots , a_n^{(M)} ),
\end{equation}
where
$ a_i^{(M)}= \sqrt{{(a_i^{(X)})}^2+{(a_i^{(Y)})}^2+{(a_i^{(Z)})}^2 }. $

In summary, the gait signal after transformation will be presented in 3 dimensions as described above.
\begin{equation}
\setlength{\arraycolsep}{2pt}
\vec{A}=\begin{bmatrix}
\vec{a}^{(Z)} & \vec{a}^{(XY)} & \vec{a}^{(M)}
\end{bmatrix}=  \begin{bmatrix}
a_1^{(Z)} & a_1^{(XY)} & a_1^{(M)} \\ \vdots & \vdots & \vdots \\ a_i^{(Z) } & a_i^{(XY)} & a_i^{(M)} \\ \vdots & \vdots & \vdots \\ a_n^{(Z)} & a_n^{(XY) } &  a_n^{(M)}
\end{bmatrix}.
\end{equation}

Since each acceleration sample is always transformed into the Earth coordinate system according to the current orientation of the mobile determined as soon as the acceleration value is returned, it is more robust than the solution proposed in \cite{Hoang13}, in that all of the signals are likely to be transformed according to the initial orientation of the mobile, which is predetermined before the user starts to walk.

%% file: p4-proposedmethod.tex
\section{\uppercase{Gait Recognition Model Construction}} \label{sec:gaitrecognition}
In this section, we propose a novel gait verification system using statistical analysis and a supervised learning, working effectively on orientation-independent gait signals obtained by using the method presented in the previous section. Our system follows the traditional pattern recognition process consisting of mandatory steps such as data preprocessing, segmentation, feature extraction and classification.
\subsection{Data Preprocessing}\label{sec:preprocessing}
\subsubsection{Linear Interpolation}\label{sec:interpolation}

\noindent As the mobile sensor is a power saving device which is simpler than standalone sensors, the sampling rate is not always stable. The time interval between two consecutive returned samples is not identical. First, we apply linear interpolation to the acquired acceleration samples to achieve gait signals at a fixed sampling rate.

 Moreover, due to the design of the mobile operating system (e.g., Android OS), the triplet of acceleration sample, orientation sample and gravity sample is not yielded simultaneously. Meanwhile in the proposed solution to handle the disorientation issue, it is required that this triplet need to be yielded concurrently. Therefore, we additionally apply the same linear interpolation to the obtained orientation samples and gravity samples. The timestamp of the interpolated acceleration samples is used as the reference axis to determine the approximate orientation vector and gravity vector yielded at the same time as the acceleration vector.

\subsubsection{Noise Elimination}\label{sec:noiseelimination}
\begin{figure}[!h]
\vspace{-0.2cm}
\centering
\includegraphics[width=7.5cm]{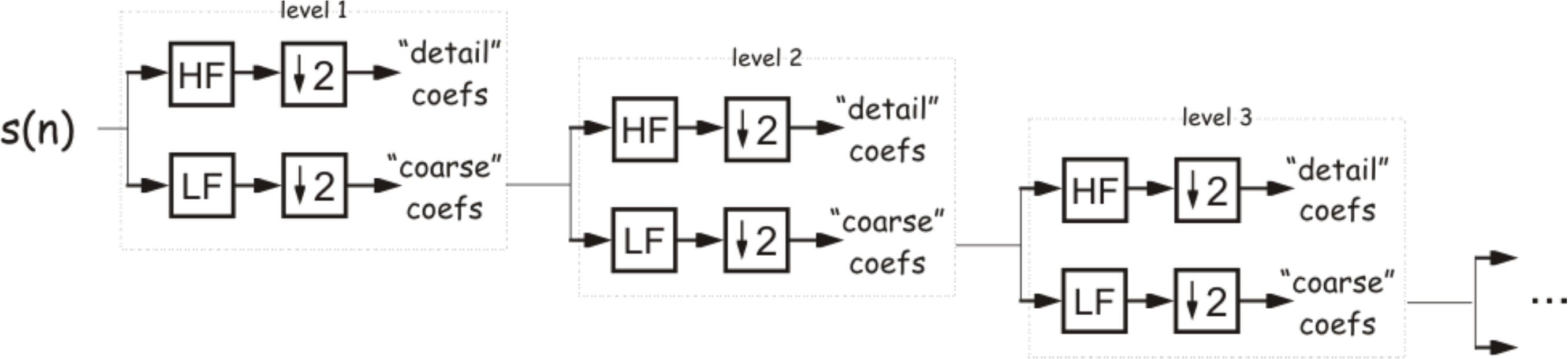}
\caption{Multi-level wavelet decomposition.}
\label{fig:noiseelimination}
\vspace{-0.1cm}
\end{figure}

\noindent Gait signals captured by the mobile accelerometer inevitably contain much noise. This can be due to the misplacement error (Figure \ref{fig:disorientation}b), the quality of the sensors or bumps on the ground while walking, the difference in footwear, etc. We apply a multi-level wavelet decomposition and reconstruction method to remove the noise components in the signal. Technically, the detailed coefficients (obtained by HF filter as in Figure \ref{fig:noiseelimination} ) are set to 0s at all decomposition levels. The signal reconstruction process involves combining the detailed coefficients of zero with the coarse coefficients from the lowest level until the level 0 is achieved. Specifically, in this study, we apply the Daubechies orthogonal wavelet (Db6) \cite{Mallat89,Daubechies92} with the decomposition at level 2 to mitigate the noise caused by the acquisition environment.

\subsection{Gait Pattern Extraction}\label{sec:gaitpatternextraction}
\subsubsection{Gait Cycle Based Segmentation}\label{sec:segmentation}
\noindent Segmentation is the most important process which could directly affect the quality of the extracted gait patterns. It can be easily seen that walking is a cyclic activity so that the gait signal should be segmented into gait cycles instead of fixed-length patterns.
\begin{figure}[!h]
\vspace{-0.2cm}
\centering
\includegraphics[width=7.5cm]{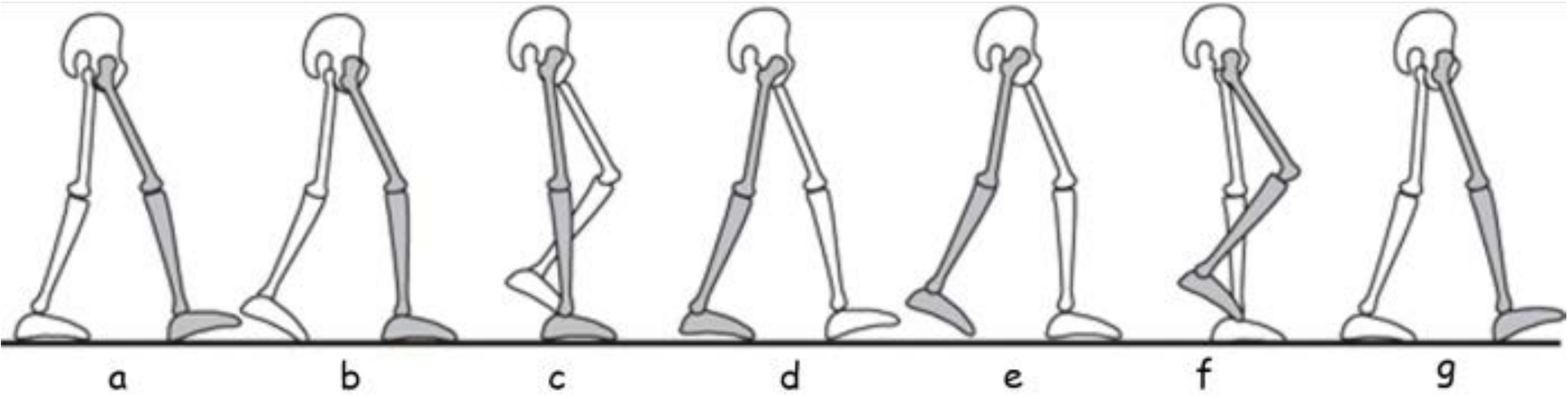}
\caption{Illustration of a gait cycle.}
\label{fig:gaitcycleillustration}
\vspace{-0.1cm}
\end{figure}

\begin{figure*}
\vspace{-0.2cm}
\centering
\subfloat[$ $]{\label{fig:gaitcyclesegmentationa}\includegraphics[width= 7cm]{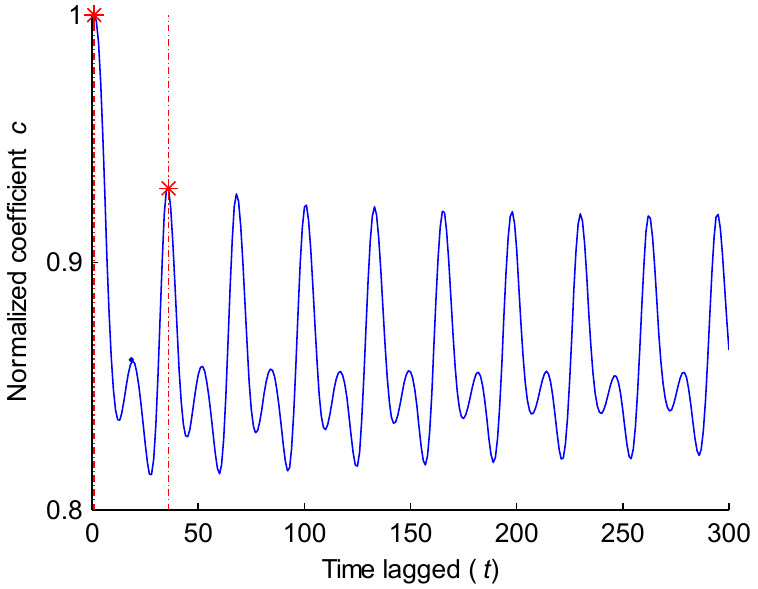}}\hfil
\subfloat[$ $]{\label{fig:gaitcyclesegmentationb}\includegraphics[width=7cm]{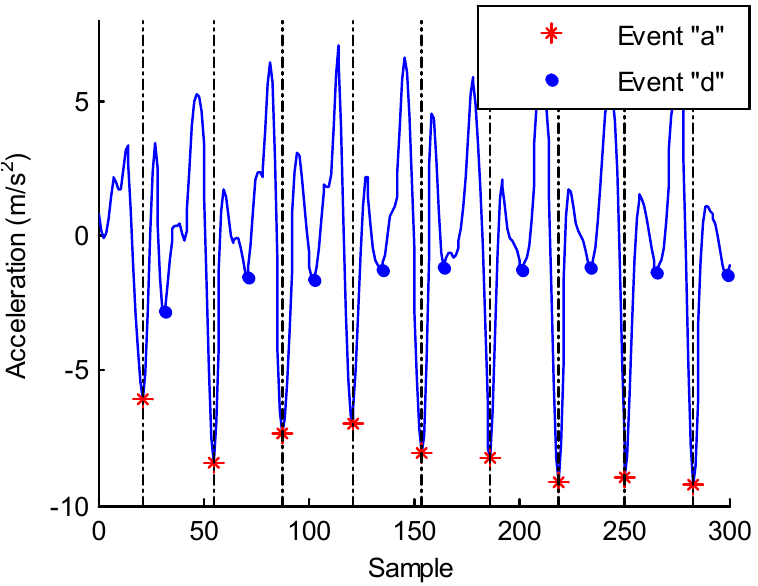}}\hfil
\caption{(a) Auto-correlation coefficients , (b) Detected marking points in $ Z- $signal.}
\vspace{-0.1cm}
\end{figure*}

Gait cycle is commonly defined as the time interval between two successive occurrences of one of the repetitive events when walking \cite{Whittle03}. Particularly, a gait cycle can start with initial contact of the heel and continue until the same heel contacts the ground again. We assume that the mobile device is placed at the same side with the leg which is going to contact the ground as in the phase ``a'' or ``g'' in the Figure \ref{fig:gaitcycleillustration}, for example the right leg. So, at the time the heel touches the ground, the association of the ground reaction force and the inertial force together will act on the right leg, which makes the acceleration value of the transformed $ Z- $dimension signal sensed by the accelerometer strongly change and form negative peaks (illustrated as star points in the Figure \ref{fig:gaitcyclesegmentationb}). They are recognized as the starting points of the gait cycles. Note that when the event ``d'' happens (e.g., the left heel touches the ground), the accelerometer also generates negative peaks, similar to the ``a'' event. However, since the device is placed at the right leg which is opposite to the left, the accelerometer only senses insignificant forces acting on the right leg in this case. Therefore, the magnitude of peaks generated by ``d'' events (Figure \ref{fig:gaitcyclesegmentationb}, circle points) is not high as those generated by ``a'' or ``g''.
The objective of the segmentation step is to divide the signal into separate gait cycles. So, it is required to determine peaks which are generated by the event of ``a'' or ``g'' in the Figure \ref{fig:gaitcycleillustration}. First of all, we determine the position of all of the negative peaks in the $ Z- $dimension gait signal $ a^{(Z)} $ of length $ n $. Let 
\begin{equation}
\Pi=\{i_j |  a_{i_j-1}^{(Z)}>a_{i_j}^{(Z)}, a_{i_j+1}^{(Z)}>a_{i_j}^{(Z)},i_j \in {1 \hdots n}\}   
\end{equation}
be the set of index of these peaks with the order preserved. Assuming that $ |\Pi| $ is always larger than 1 given a gait signal of a walking session, we filter the starting points of the gait cycles in $ \Pi $ based on two criteria.

The first criterion is based on the magnitude of the peaks. We eliminate the noisy peaks whose values are higher than a threshold $ \delta $ determined by the mean and standard deviation of all of the peaks in $ \Pi $.
\begin{equation}
\delta=\mu_\Pi - \tau\sigma_{\Pi},  	
\end{equation}
where
\begin{equation}
\begin{aligned}
\mu_\Pi& =\frac{1}{|\Pi|} \sum_{i\in \Pi}^{}{a_i^{(Z)} }, \\ 
\sigma_\Pi&=\sqrt{\frac{1}{(|\Pi|-1)} \sum_{i\in \Pi}{}{a_i^{(Z)}-\mu_\Pi}} ,\\
\end{aligned}
\end{equation}
$ \tau $ is a user-defined parameter.

The second is based on the correct positions of the gait cycle's starting points. While the distance between starting points of the gait cycle is assumed to fluctuate around a constant range in other studies \cite{Derawi101,Derawi102}, we observed that such range does not cover all possible cases since the walking speed of different individuals varies significantly. Instead, we estimate the length of the gait cycle according to the characteristics of each signal.

To determine the periodicity of the gait signal, we calculate the autocorrelation coefficients $ c_t $ ($ 0\le t< n $) of the $ Z- $dimension signal by
\begin{equation}
c_t=\frac{N}{N-t}\times\frac{\sum_{i=1}^{N-t}{{a_i}}^{(Z)}{a_{i+t}}^{(Z)}}{\sum_{i=1}^{N}{({a_i}^{(Z)})}^2}.
\end{equation}

The moving average algorithm is then applied to smooth these coefficients. Let us assume that $ c_i $ and $ c_j $ are the $ 1^{st}  $ and $ 2^{nd} $ peaks autocorrelation coefficients, respectively, as depicted by two stars in the Figure \ref{fig:gaitcyclesegmentationa}. Then, the length of a gait cycle can be approximated by 
\begin{equation}
\Delta =j.
\end{equation}

According to the two criteria of magnitude and position, we determine the peaks representing the starting points of the gait cycles. Let 
$ \Omega $ be the set of these peaks with the order of indices preserved. Then, $ \Omega $ will be given by
\begin{equation}
\resizebox{0.48\textwidth}{!}{$
\Omega=\{i_j | a_{i_j}^{(Z)}< \delta ,(\exists i_k \in \Pi ,k>j,\Delta-\epsilon \le i_k -i_j \le \Delta + \epsilon),i_j \in \Pi\},
$}
\end{equation}
where $ \epsilon >0 $ is the user-defined parameter.
Assuming that there are $ k-1 $ where $ k>1 $ gait cycles in the given gait signal, hence, $ |\Omega|=k $. We separate the signal into $ k-1 $ distinct segments, with each $ \vec{S}_i $ consisting of a full gait cycle
\begin{equation}
\vec{S}_i  = \begin{bmatrix}
a_{\Omega_i}^{(Z)} & a_{\Omega_i}^{(XY)} & a_{\Omega_i}^{(M)}\\ \vdots & \vdots & \vdots \\a_{\Omega_{i+1}}^{(Z)} & a_{\Omega_{i+1}}^{(XY)} & a_{\Omega_{i+1}}^{(M)}
\end{bmatrix}.
\end{equation}

\begin{table*}
\caption{List of gait features extracted in time and frequency domains}\label{tab:extractedFeatureList}
\begin{tabular}{|c|}
  \hline
\textbf{  Time domain features}\\
  \hline
\makecell{Mean of the max/min value in each $ \vec{s}_u^{(D)} $ in $ \vec{p}^{(D)} $ where $ D = Z, XY M$; average absolute difference; root mean\\ square; standard deviation; waveform length; 10-bin histogram distribution; average length of $ \vec{s}_u^{(D)} $ in $ \vec{p}^{(D)} $}\\
  \hline
 \textbf{ Frequency domain feature}s\\
  \hline
  Magnitudes of first 40 FFT coefficients; first 40 DCT coefficients.\\
  \hline
  \end{tabular}
\end{table*}
\subsubsection{Pattern Extraction}\label{sec:patternextraction}
We form gait patterns by concatenating the separate one-gait-cycle segments extracted in the previous section. Each gait pattern would contain $ n_s $ consecutive segments and 50\% of them overlaps with the previous one.
Let $ \begin{bmatrix}
\vec{s}_u^{(Z)}& \vec{s}_u^{(XY)} & \vec{s}_u^{(M)} 
\end{bmatrix} $ be a segment consisting of a gait cycle, where $ \vec{s}_u^{(Z)}= \begin{bmatrix}
a_{u1}^{(Z)} \hdots a_{un_u}^{(Z)}
\end{bmatrix}^\top, \hspace{2pt} \vec{s}_u^{(XY)}=\begin{bmatrix}
a_{u1}^{(XY)} \hdots a_{un_u}^{(XY)}
\end{bmatrix}^\top,  \hspace{2pt} \vec{s}_u^{(M)}=\begin{bmatrix}
a_{u1}^{(M)} \hdots a_{un_u}^{(M)}
\end{bmatrix}^\top $. Let \\
\resizebox{0.475\textwidth}{!}
{$ \setlength{\arraycolsep}{1pt} \setcounter{MaxMatrixCols}{15} 
\vec{p}^{(Z)}=\begin{bmatrix}
a_{11}^{(Z)} & \hdots & a_{1n_1}^{(Z)} & a_{21}^{(Z)} & \hdots &  a_{2n_2}^{(Z)} & \hdots & a_{u1}^{(Z)} & \hdots & a_{un_{u}}^{(Z)} & \hdots & a_{s1}^{(Z)} & \hdots & a_{sn_s}^{(Z)}
\end{bmatrix}^\top, $}\\ 
\resizebox{0.49\textwidth}{!}
{$  \setlength{\arraycolsep}{1pt} \setcounter{MaxMatrixCols}{15}
\vec{p}^{(XY)}=\begin{bmatrix}
a_{11}^{(XY)} & \hdots & a_{1n_1}^{(XY)} & a_{21}^{(XY)} & \hdots & a_{2n_2}^{(XY)} & \hdots & a_{u1}^{(XY)} & \hdots & a_{un_u}^{(XY)} & \hdots & a_{s1}^{(XY)} & \hdots & a_{sn_s}^{(XY)}
\end{bmatrix}^\top, $}\\ \resizebox{0.48\textwidth}{!}
{$ \setlength{\arraycolsep}{1pt} \setcounter{MaxMatrixCols}{15}
\vec{p}^{(M)}= \begin{bmatrix}
a_{11}^{M)} & \hdots & a_{1n_1}^{(M)} & a_{21}^{(M)} & \hdots & a_{2n_2}^{(M)} & \hdots & a_{u1}^{(M)} & \hdots & a_{un_u}^{(M)} & \ \hdots & a_{s1}^{(M)} & \hdots & a_{sn_s}^{(M)}
\end{bmatrix}^\top. $}

Then, a gait pattern $ \vec{P} $ is defined by
\begin{equation}\label{eq:extractedPattern}
\setlength{\arraycolsep}{2pt}
\vec{P}= \begin{bmatrix}
\vec{p}^{(Z)} & \vec{p}^{(XY)} & \vec{p}^{(M)}
\end{bmatrix}= \begin{bmatrix}
a_{11}^{(Z)}& a_{11}^{(XY)} & a_{11}^{(M)}\\ \vdots & \vdots & \vdots \\ a_{un_u}^{(Z)}& a_{un_u}^{(XY)} & a_{un_u}^{(M)}\\ \vdots & \vdots & \vdots \\ a_{sn_s}^{(Z)}& a_{sn_s}^{(XY)}& a_{sn_s}^{(M)}
\end{bmatrix}.
\end{equation}

\subsection{Gait model construction}\label{sec:gaitmodelconstruction}
\subsubsection{Feature extraction}\label{sec:featureextraction}
We extract the features on both time and frequency domains as used in \cite{Hoang13} for each gait pattern $ \vec{P} $. The list of extracted features is briefly summarized in the Table \ref{tab:extractedFeatureList}. Note that all of the features in the time and frequency domains are extracted for the 3 dimensions of the gait pattern (viz. $ \vec{p}^{(Z)}, \vec{p}^{(XY)}, \vec{p}^{(M)} $ in (\ref{eq:extractedPattern})), except for the ``average length of  $ \vec{s}_u^{(D)} $ in $ \vec{p}^{(D)} $'' feature since its value is identical in all 3 dimensions. All of the extracted features are concatenated to form the final feature vector for a gait pattern.

\subsubsection{Feature vector dimension reduction}\label{sec:dimensionreduction}
\noindent Since we expect the system to run directly on the mobile phone with limited computational resources, it is necessary to reduce the dimension of the extracted feature vectors to lighten the complexity of the gait model built by using the machine learning algorithms. Thus, we adopt the Principle Component Analysis  (PCA) to reduce the number of dimensions while maintaining the discriminability of the feature vectors. 

Let us assume that the number of users is denoted as $ N $. The number of feature vectors extracted from all of the gait patterns for each user is $ M $. According to the feature extraction phase, the length of each feature vector is $ n_f= 289 $. The $ j^{th} (j =1 \hdots M) $ feature vector of the user $ i (i = 1 \hdots N) $ is denoted as
\begin{equation}
\vec{v}_j^{(i)}=\begin{bmatrix}
f_{j,1}^{(i)},\hdots,f_{j,k}^{(i)},\hdots,f_{j,n_f}^{(i)} 
\end{bmatrix},
\end{equation}
where $ f_{j,k}^{(i)} $ is the $ k^{th}$ feature component of $ \vec{v}_j^{(i)} $. The matrix of feature vectors of all users can be formed as
\begin{equation}
\noindent
\setlength{\arraycolsep}{2pt}
\resizebox{0.48\textwidth}{!}
{$
\vec{F}^\top=\begin{bmatrix}
\vec{v}_1^{(1)}\ \\ \vdots \\ \vec{v}_{j}^{(i)} \\ \vdots \\\vec{v}_M^{(N)}
\end{bmatrix} = \begin{bmatrix}
f_{1,1}^{(1)} & f_{1,k}^{(1)} & f_{1,n_f}^{(1)} \\ \vdots & \vdots &\vdots \\ f_{j,1}^{(i)} & f_{j,k}^{(i)} & f_{j,n_f}^{(i) }\\ \vdots & \vdots & \vdots \\ f_{M,1}^{(N)} & f_{M,k}^{(N)} & f_{M,n_f}^{(N) }
\end{bmatrix} = \begin{bmatrix}
\vec{v}_1\\ \vdots \\ \vec{v}_t \\ \vdots \\ \vec{v}_{MN} 
\end{bmatrix} \in \mathbb{R}^{MN\times n_f}.
$}	
\end{equation}
Then, a covariance matrix of $ \vec{F} $ can be calculated by 
\begin{equation}
\Sigma = \frac{1}{MN}\sum_{i=1}^{MN}{(\vec{v}_i-\bar{\vec{v}})(\vec{v}_i-\bar{\vec{v}})^\top} \in \mathbb{R}^{n_f\times n_f}.
\end{equation}
Let $ \overrightarrow{\lambda}= (\lambda_1,\hdots,\lambda_i,\hdots,\lambda_{n_f}) $ and $ \vec{u}_i $ be eigenvalues and eigenvectors obtained from the $ \Sigma $, respectively. All eigenvalues $ \lambda_i $  of $ \Sigma $ are sorted in descending order in which the higher the eigenvalues are, the more important they are. Assuming that $ \lambda_i<\lambda_{i-1} $, to reduce the number of dimensions of the original feature vector from $ n_f $ to $ k $, $ k  $ eigenvectors are taken according to the order of the eigenvalues
\begin{equation}
\vec{U} = \begin{bmatrix}
\vec{u}_1 & \hdots & \vec{u}_i & \hdots & \vec{u}_k
\end{bmatrix} \in \mathbb{R}^{n_f \times k}.
\end{equation}
The dimension-reduced matrix of feature vectors can be calculated by
\begin{equation}
\vec{\hat{F}}^\top= \vec{F}^\top \vec{U}.
\end{equation}

\subsubsection{Gait recognition model for verification and identification}\label{sec:classification}
We adopt two schemes, namely feature vector matching and supervised learning, for both identification and verification. In the former scheme, the feature vectors extracted after using PCA are stored in the mobile storage, which will be used for user identification or verification. In the latter scheme, we apply Support Vector Machine (SVM) with a linear kernel to build a gait model for each user. An open library tool, libsvm \cite{Chang11}, is used in this study for SVM-based gait model construction and evaluation.

%% file: p5-evaluation.tex
\section{Experiments}\label{sec:experiment}
\subsection{Dataset description}\label{dataset}
\noindent We use the dataset which is an extended version of the one used in \cite{Hoang13} for experimental evaluation in this study. We would like to briefly describe the original and extensions of this dataset and refer the readers to the original work of the authors for more details. The dataset consists of gait signals of 38 subjects captured by using a Google Nexus One mobile phone. The device is put freely inside the front trouser pocket and the sampling rate of integrated sensors is set to 27Hz. Besides data described in the original work, the authors further provide other sensor data which are collected along with acceleration data during  gait sensing period in the extended version. List of particular sensors activated in this phase is summarized in the Table \ref{tab:sensorspecification}.
\begin{table}[b]
\caption{List of physical and virtual sensors activated during the gait capture process }\label{tab:sensorspecification}
\resizebox{\columnwidth}{!}{%
\begin{tabular}{|c | c | c |}
  \hline
  Sensor Name & Model name & Sampling rate\\
  \hline
  Magnetic field sensor & AK 8973 & 25 Hz \\
  Accelerometer & BMA 150 &  25 Hz\\
  Orientation sensor & Virtual & 25 Hz\\
  Gravity sensor & Virtual & 25 Hz\\
  Linear acceleration sensor & Virtual & 25 Hz\\
  Rotation vector sensor & Virtual & 25 Hz\\
  
  \hline
\end{tabular}
}
\end{table}
\begin{table*}
\caption{ The configuration differences in between the original study and this experiment }\label{tab:experimentsetup}
\begin{tabular}{|c | c | c | c | c | c | c| c | c|}
  \hline
  \multirow{2}{*}{Method}& \multicolumn{4}{c|}{Original setup} & \multicolumn{4}{c|}{This experiment} \\
  \hhline{~--------}
	& Axes & $ \# $ Subject & Position & SR (\%) & Axes & $ \# $ Subject & Position & SR \\
  \hline
  Rong et al.& $ X,Y,Z $ & 38 & Ankle & 250 & $ Z,XY,M $ & 38 & Front pocket & 27 \\
  Gafurov et al. & $ Z $ & 30 & Ankle & 100 & $ Z $ & 38 & Front pocket & 27 \\
  Derawi et al. & $ M$ & 60 & Hip & 100 & $ M $ & 38 & Front pocket & 27 \\
  \hline
\end{tabular}
\end{table*}

\subsection{Experimental configuration}\label{sec:experimentconfig}
\noindent Since the sampling rate of the sensor is low (27 Hz), making the number of samples in a one-gait-cycle segment small, we form each gait pattern extracted in Section \ref{sec:patternextraction} by concatenating $ n_s = 4 $ consecutive segments, in order to feasibly extract enough features in the time and frequency domains. In total, around 10226 gait patterns are extracted from the dataset. Moreover, the length of the feature vectors after applying PCA is selected to be equal to $ n' $ such that the first $ n'  $ eigenvectors capture at least 99.5\% of the total variance. According to the dataset used in this study, $ n' $ is equal to 42.

We re-implement several state-of-the-art gait recognition systems on the dataset used in this experiment (\cite{Rong07,Gafurov10,Derawi102}) in order to not only evaluate the effectiveness of the solution proposed to handle the disorientation problem, but also compare with our proposed recognition schemes. The effectiveness of gait recognition systems is evaluated under two aspects: identification and verification capabilities. Note that in the comparison of the verification and identification capabilities among the studies, the disorientation problem is not taken into account so that all of the works will be evaluated on the orientation-independent gait signals (referred to as the transformed dataset). Furthermore, as Rong et al. used the gait signals of the $ X- $ and $ Y- $ dimensions which are not available in the transformed dataset, we replace them with the gait signals of the $ XY- $ and $ M- $ dimensions to make sure that the number of dimensions is consistent to that in the original study. In Gafurov et al.'s method, the authors experimented on the gait signals of different dimensions and achieved various results. Based on the availability of dimensions in the transformed dataset, we select the gait signal of the $ Z- $dimension (referred to as the up-down direction in the original) as the standard for evaluation and comparison. Table \ref{tab:experimentsetup} shows the difference in the configuration settings in between the original studies and this experiment.

\begin{figure*}
\vspace{-0.2cm}
\centering
\subfloat[PCA]{\label{fig:rocdifferentportiondataa}\includegraphics[width= 7.5cm]{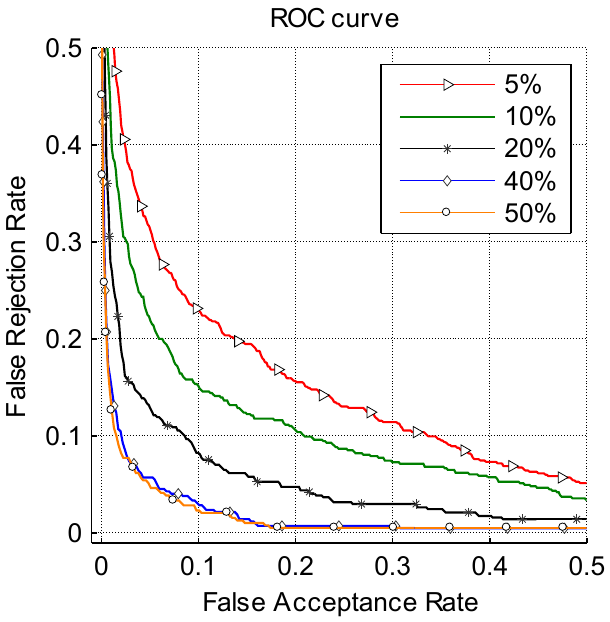}}\hfil
\subfloat[PCA+SVM] {\label{fig:rocdifferentportiondatab}\includegraphics[width=7.5cm]{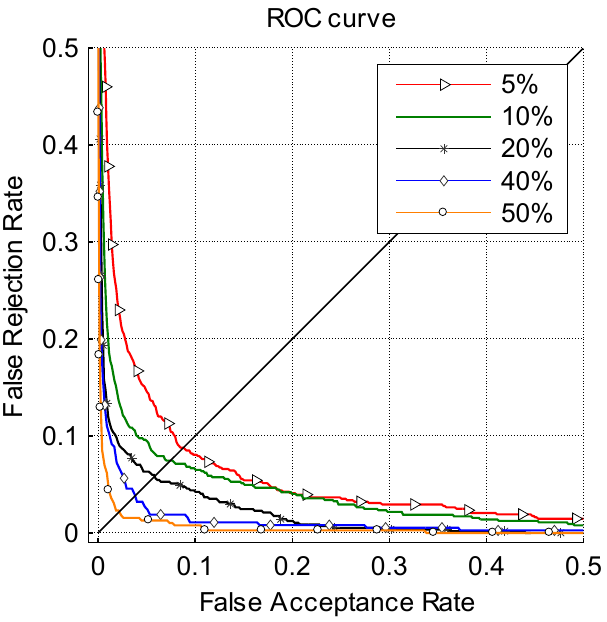}}\hfil
\caption{ROC curves of the proposed system using two verification schemes according to different portions of training data.}
\label{fig:rocdifferentportiondata}
\vspace{-0.1cm}
\end{figure*}

\begin{figure*}
\vspace{-0.2cm}
\centering
\subfloat[Session-based verification] {\label{fig:roccomparisona}\includegraphics[width= 7.5cm]{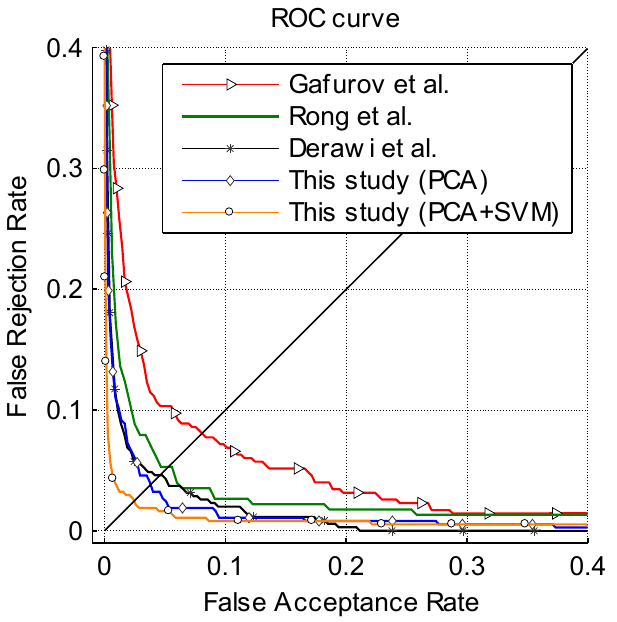}}\hfil
\subfloat[Pattern-based verification] {\label{fig:roccomparisonb}\includegraphics[width=7.5cm]{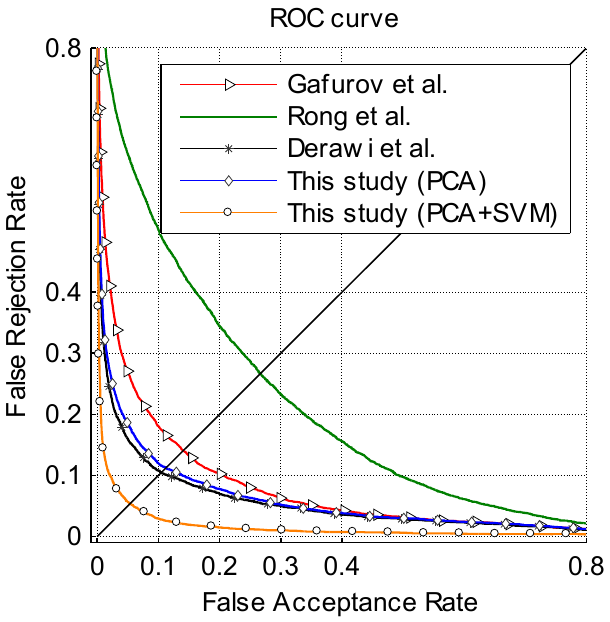}}\hfil
\caption{The ROC curves of the proposed method and other studies which are re-implemented and evaluated according to the configuration in Table \ref{tab:experimentsetup}}.
\label{fig:roccomparison}
\vspace{-0.1cm}
\end{figure*}

\begin{figure*}
\vspace{-0.2cm}
\centering
\subfloat[Rong et al.] {\label{fig:rocimpactRong}\includegraphics[width= 3.86cm]{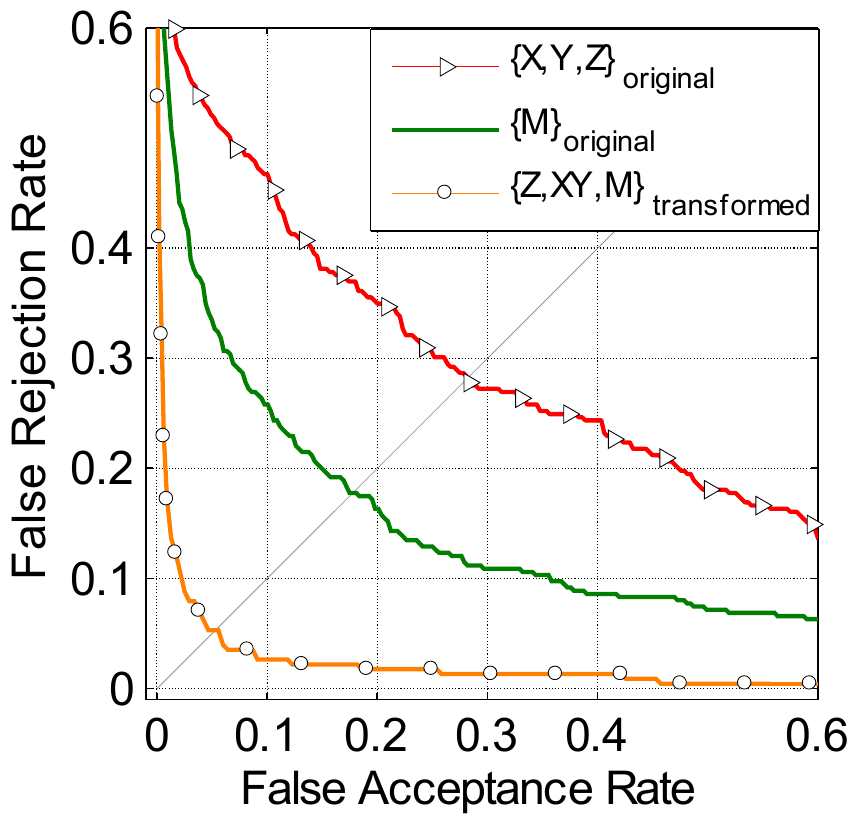}}\hfil
\subfloat[Gafurov et al.] {\label{fig:rocimpactGafurov}\includegraphics[width= 3.75cm]{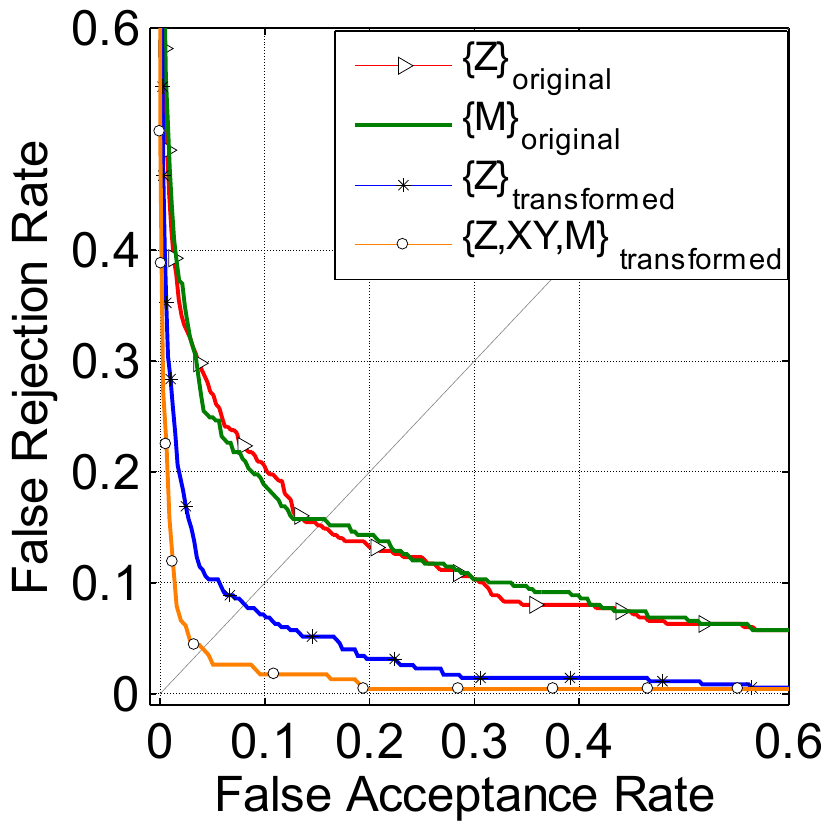}}\hfil
\subfloat[Derawi et al.] {\label{fig:rocimpactDerawi}\includegraphics[width=  3.75cm]{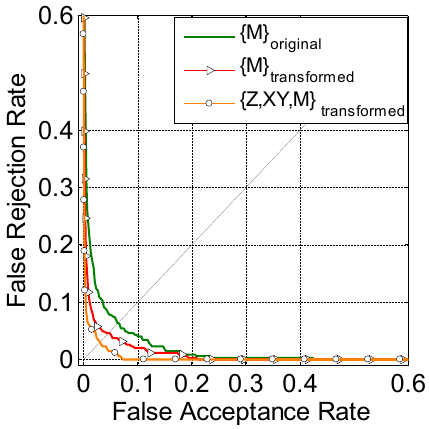}}\hfil
\subfloat[The proposed method] {\label{fig:rocimpactProposed}\includegraphics[width= 3.75cm]{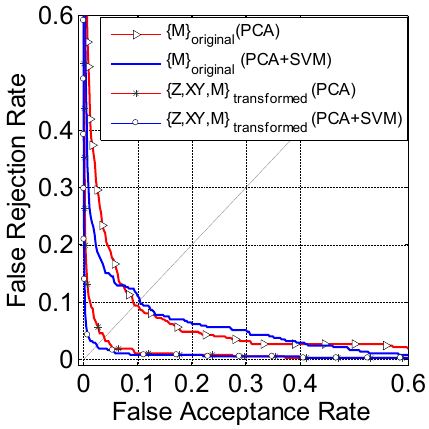}}\hfil

\caption{The impacts of the disorientation error on the error rates of the gait verification systems.}
\label{fig:rocimpact}
\vspace{-0.1cm}
\end{figure*}

\subsection{Verification results}\label{sec:verificationresult}
\begin{table*}
\caption{ The error rates of session-based and pattern-based gait verification methods}\label{tab:verification}
\begin{tabular}{|c | c | c | c | c | c |}
  \hline
  \multirow{3}{*}{Method}& \multicolumn{3}{c|}{Session-based} & \multicolumn{2}{c|}{Pattern-based} \\
  \hhline{~-----}
	& \makecell{EER(\%) \\ (\textit{original})} & EER (\%) & \makecell{FRR (\%) \\ (\textit{at} FAR = 1\%) }& EER (\%) & \makecell{FRR (\%) \\ (\textit{at} FAR = 1\%)}\\
  \hline
  Rong et al.& 5.6 & 5.28 & 16.47 & 26.67 & 84.27 \\
  Gafurov et al. & $ 2.2 - 23.6 $ & 8.07 & 28.43 & 14.11 & 52.37 \\
  Derawi et al. & 5.7 & 4.59 & 10.71 & 10.49 & 31.86 \\
  Proposed method (PCA) & -- & 3.83 & 10.75 & 11.23 & 35.03 \\
  Proposed method (PCA+SVM) & -- & \textbf{2.45} & \textbf{3.75} & \textbf{5.35} & \textbf{14.38}\\
  \hline
\end{tabular}
\end{table*}
\noindent We utilize Receiver Operating Characteristic (ROC) curves to illustrate the performance of the proposed system in the aspect of verification. Firstly, we experiment with different portions of the training data and testing data, ranging from  $ k = 5\%  $ to $ 50\% $, in order to determine the influence of the number of training data on the effectiveness of the proposed schemes. Note that we apply cross-verification to overall evaluate the performance of the proposed method. Participants will be considered as the genuine user in turn. Specifically for each user $ i $, in the PCA approach, we store randomly $ k $ gait patterns of the user $ i $ as the training data. The remaining patterns of the user $ i $ will be utilized for testing the false rejection rate (FRR), and the patterns of all other users $ j (j \ne i) $ will be used for testing false acceptance rate (FAR). Similarly in the PCA+SVM approach, $ k $ genuine gait patterns of the user $ i $ and $ k $ patterns of each other users will be used to construct the gait model for the user $ i $, while remaining data will be used for evaluating the error rates of the constructed model.

Figure \ref{fig:rocdifferentportiondata} depicts the error rates of the proposed method using the PCA and PCA+SVM schemes with different proportions of training data and testing data. As expected, a higher proportion of the training data yield a lower error rate. Moreover, we can see that applying a supervised learning (PCA+SVM) technique can help to enhance the accuracy of the system. The overall error rate achieved using the PCA+SVM scheme is lower than that when only using the PCA scheme. 
Next, we compare the proposed system with those of other studies which are re-implemented and evaluated according to the new configuration settings. The verification performance of all of these studies is investigated on two testing scenarios: Firstly, we consider each walking session as a testing trial, which is commonly used in comparing studies. Unlike in these studies, we apply majority voting to our schemes to validate each walking session. That means the user is verified if a larger portion of the gait patterns extracted in the session is recognized as being authentic. Figure \ref{fig:roccomparisona} depicts the error rates of the proposed methods and other studies in this scenario. 

From our viewpoint, session-based verification might require a lot of time and efforts from the user since he/she has to continuously walk for a long distance, in order to collect enough data to be verified. The verification process can be performed faster and more constantly if the system can immediately verify the user only using the gait pattern instead of having to wait until the walking session finishes. Therefore, we additionally investigate the performance of the methods in all of the studies when the separate gait patterns are treated as independent testing trials. As depicted in Figure \ref{fig:roccomparisonb}, we can see that the error rates of all of the approaches are higher than those in the session-based scenario. Especially in Rong et al.'s method, the error rate is significantly increased, since the method of gait pattern extraction employed strongly relies on the whole walking session data. The error rate of our method using the PCA+SVM scheme in this scenario is approximately 5.35\%, which can help to reduce the time and effort needed to perform the verification task. Table \ref{tab:verification} summarizes the EERs achieved in all of the studies according to two scenarios. It can be seen that in the session-based scenario, the achieved EERs with the methods proposed in the other studies after evaluating them in the transformed dataset are similar to original values. This reflects that handling the disorientation problem is mandatory in order to maintain the effectiveness of gait recognition systems because this problem might result in a significant increase of the error rates of the systems. This impact will be clearly shown  in the Section \ref{sec:impactdisorientation}.

\subsection{Identification results}\label{sec:identificationresult}
\begin{table}
\caption{ The error rates of session-based and pattern-based gait identification methods}\label{tab:identification}
\resizebox{\columnwidth}{!}
{%
\begin{tabular}{|c | c | c |}
  \hline
  \multirow{2}{*}{Method}& \multicolumn{2}{c|}{Accuracy rate (\%)} \\
  \hhline{~--}
	& Session-based & Pattern-based \\
  \hline
  Rong et al. (+kNN) & 93.12 & 64.82 \\
  Gafurov et al. (+kNN) & 87.68 & 76.55 \\
  Derawi et al. (+kNN)& 93.41 & 88.09\\
  Proposed method (PCA+kNN) & 96.56 & 85.48 \\
  Proposed method (PCA+SVM) & \textbf{99.14} & \textbf{94.93}\\
  \hline
\end{tabular}
}
\end{table}
We also investigate the identification capability of the proposed method and other studies according to the two evaluation scenarios described above. The same 1-nearest neighbor algorithm is applied to all methods, except for the PCA+SVM scheme in order to measure the performance between studies. The best accuracy rate belongs to the proposed PCA+SVM scheme, with an amount of approximately 99.14\% being achieved under session-based identification aspect. Table \ref{tab:identification} shows the identification performance of the proposed method and comparing studies. Similar to the verification results, the accuracy rate of pattern-based identification is normally lower than that of session-based identification. Especially in Rong et al.'s method, the accuracy of the system strongly decreases an amount of approximately 30\%.
\subsection{The impacts of disorientation error}\label{sec:impactdisorientation}

Finally, we illustrate the impact of disorientation errors on the accuracy of the gait verification systems. As already mentioned, the instability of sensor orientation would cause the gait signals acquired in the 3 separate dimensions to be dissimilar. As can be seen in Figure \ref{fig:rocimpactRong} (the triangle line), since the authors store the gait patterns of the separate dimensions, including the  X-Template, Y-Template and Z-Template, as the reference set for individual matching, the error rate is significantly increased, because of the dissimilarity issues. This is similar to Gafurov et al.'s work (Figure \ref{fig:rocimpactGafurov}). Looking at both Figures \ref{fig:rocimpactRong}, \ref{fig:rocimpactGafurov}, it can be seen that protecting the similarity of the gait signals from the disorientation problem can help to maintain the accuracy rate of the system.
Based on our observations, the magnitude of the gait signal is orientation-independent, so it can be used to construct the gait verification system in spite of the disorientation issues. This signal was used in the original study of Derawi et al. and achieved positive results (Figure \ref{fig:rocimpactDerawi}). Therefore, we also modified the methods of Rong et al. and Gafurov et al. by only using the magnitude of the signal and found that the error rates could be enhanced. However, from our perspective, we are strongly convinced that the gait can be more distinguishable if the gait signal can be expressed in higher dimensions. Consequently, additional experiments are conducted according to the hypothesis, wherein we employ the gait signals of all dimensions, which can be obtained after overcoming the disorientation problem. As expected, the error rates achieved with the methods of all of the studies are likely to be more decreased when the gait signals in the dimensions of $ Z, XY, M $ are all used (Figure \ref{fig:rocimpact}). Therefore, we believe that overcoming the disorientation problem, in order to maintain the number of dimensions of the acquired gait signals, is mandatory to optimize the performance of gait verification and identification systems.

%% file: p6-conclusion.tex
\section{\uppercase{Conclusions}}
\label{sec:conclusion}

\noindent In this paper, we addressed the sensor disorientation problem in gait verification or identification systems which can frequently arise in reality, especially in the mobile context. A simple but effective solution taking advantages of available sensors in mobile device was proposed. A gait recognition model leveraging statistical analysis and supervised machine learning which could be used to verify or identify mobile user was also presented. The results achieved are highly promising, especially with regard to identification. They reflect the good potential of deploying a gait-based authentication to ameliorate the security on portable devices. Note that our proposed method does not aim to completely replace the existing explicit authentication schemes on mobiles, since at this moment it is infeasible to achieve a perfect security level (e.g., the zero-FAR is always achieved) of any behavioral biometric-based verification systems. However, the proposed method can be used as an additional authentication scheme, especially for applications which do not require excessively strict security levels, to enhance the usability of the device. In a future work, we would like to investigate on developing a unique gait recognition model working effectively regardless of the relative position of the mobile to its owner. A protection scheme used to secure gait template/ models stored directly in the device will also be our main further work.

\section*{\uppercase{Acknowledgements}}
This research was supported by Basic Science Research Program through the National Research Foundation of Korea (NRF) funded by the Ministry of Education (2012R1A1A2007014).

The research was also supported by 2012-18-02TD VNU--HCMC Project.